\documentclass[12pt]{amsart}%
\usepackage{amsfonts}
\usepackage{graphicx}
\usepackage{rotating}
\usepackage{pstricks, pst-node, pst-text, pst-3d}
\usepackage{amsmath}
\usepackage{amssymb}%
\usepackage{caption}
\input{amssym.def}
\theoremstyle{remark}

\newcommand{\im}{\text{\rm Im}\,}

\newcommand{\tb}{\text{\bf t}}
\newcommand{\ub}{\text{\bf u}}

\input epsf
\begin{document}
\title[L\"owner evolution and finite dimensional reductions\dots]{L\"owner evolution and finite dimensional reductions of integrable systems}
\author[M.V.~Pavlov, D.~Prokhorov, A.~Vasil'ev, and A.~Zakharov]{Maxim V. Pavlov$^{\sharp}$, Dmitri Prokhorov$^{\ddag}$,
Alexander Vasil'ev$^{\dag}$,\\ and Andrey Zakharov$^{\ddag}$}
\address{M.~V.~Pavlov: Mathematical Physics Department, Lebedev Physical Institute, Russian Academy of Sciences,
Leninskij Prospekt 53, Moscow  119991, Russia}
\email{maksmath@gmail.com}

\address{D.~Prokhorov: Department of Mechanics and Mathematics, Saratov State University,
Astrakhanskaya Str.~83, Saratov 410012, Russia}
\email{prokhorov@sgu.ru}

\address{A.~Vasil'ev: Department of Mathematics, University of Bergen, Johannes Brunsgate 12, Bergen 5008, Norway}
\email{Alexander.Vasiliev@math.uib.no}

\address{A.~Zakharov: Department of Mechanics and Mathematics, Saratov State University,
Astrakhanskaya Str.~83, Saratov 410012, Russia}
\email{zakharovam66@gmail.com}

\thanks{The author$^{\sharp}$ has been supported by the RF Government grant
\#2010-220-01-077, ag. \#11.G34.31.0005, by the grant of Presidium of RAS
\textquotedblleft Fundamental Problems of Nonlinear
Dynamics\textquotedblright\ and by the RFBR grant 14-01-00389. The author$^{\dag}$ has been  supported by the grants of the Norwegian Research Council \#204726/V30,
\#213440/BG; and EU FP7 IRSES program STREVCOMS, grant  no.
PIRSES-GA-2013-612669. The authors$^{\ddag}$ have been supported by the
Russian/Turkish grant RFBR/T\"UBITAK \#14-01-91370}

\dedicatory{Dedicated to Ludwig D.~Faddeev on the occasion of his 80th birthday}



\begin{abstract}
The L\"owner equation is known as a one-dimensional reduction of the Benney chain as well as the dispersionless KP hierarchy. We propose a reverse process showing that time splitting in the L\"owner or the L\"owner-Kufarev equation leads to some known integrable systems.
\end{abstract}
\maketitle

\section{Introduction}

One of the central problems in the theory of integrable systems is their finite-dimensional reductions. We start with the dispersionless Kadomtsev--Petviashvili (dKP) hierarchy
as an illistration. Let $\lambda(z,\tb)$ be a meromorphic function in variable $z$ and depending on an infinite family of generalized times $\tb=(t_0=x, t_1,\dots,t_n\dots)$ with
the expansion 
\[
\lambda(z,\tb)=z+\sum\limits_{n=0}^{\infty}\frac{A^n(\tb)}{z^{n+1}},
\]
about infinity (here $A^n$ means index instead of exponent). The Poisson structure $\{\cdot\, , \cdot\}$ is defined by
\[
\{F,G\}=\frac{\partial F}{\partial z}\frac{\partial G}{\partial x}-\frac{\partial F}{\partial x}\frac{\partial G}{\partial z}.
\]
Then the dKP hierarchy (see \cite{KodamaG89}) is an infinite number of commuting flows
\begin{equation}\label{dKPE}
\frac{\partial \lambda}{\partial t_{n}}=\{\mathcal L_{n+1},\lambda\},\quad n\geq 0,
\end{equation}
where $\mathcal L_n=\frac{1}{n}(\lambda^n)_{\geq 0}$, $n=1,2,\dots$, denotes the polynomial part of $\lambda^n$. In particular, $(\lambda)_{\geq 0}=z$, $(\lambda^2)_{\geq 0}=z^2+2A^0$, $(\lambda^3)_{\geq 0}=z^3+3zA^0+3A^1$, etc. For $n=2$ and $s=t_1$, the second equation in the hierarchy is equivalent to
\begin{equation}\label{BE}
A_s^n+A_x^{n+1}+nA^{n-1}A_x^0=0,
\end{equation}
known as the Benney moment equation.
Benney~\cite{Benney} investigated long non-linear waves
propagating on a free surface showing that the governing equations
have an infinite number of conservation laws. 

The compatibility condition for \eqref{dKPE} is
\begin{equation}\label{compa}
\frac{\partial \mathcal L_m}{\partial t_n}-\frac{\partial \mathcal L_n}{\partial t_m}+\{\mathcal L_m,\mathcal L_n\}=0,
\end{equation}
which means that the flows   \eqref{dKPE} commute. In particular, for $m=3$, $n=2$,  $y=t_2$ and $u:=A^0$ we arrive at the equation
\begin{equation}\label{KZ}
u_{ss}+\frac{\partial}{\partial x}(u_{y}+uu_x)=0, 
\end{equation}
known as the dKP equation or the Zabolotskaya--Khokhlov equation~\cite{ZabKhokh}.

A finite-dimensional reduction suggests that the function $\lambda$ depends on the generalized times $\tb$ via a finite number of functions $u_k=u_k(\tb)$, i.e., 
$$\lambda(z, \tb)=\lambda(z, \ub(\tb))=\lambda(z, u_1(\tb),\dots,  u_N(\tb)).$$ A well-known polynomial reduction was proposed by Zakharov~\cite{Zakharov}.

 Kodama and Gibbons~\cite{KodamaG89}  realized that
dKP equation possesses infinitely many multi
component two dimensional reductions. They
presented several examples and  the dependence
$\lambda$ was found in these particular cases.
 They considered the vector-function $\ub$ satisfying a system of hydrodynamic type
\[
\frac{\partial \ub}{\partial t_n}=\xi_n(\ub)\frac{\partial \ub}{\partial x}, \quad n>1.
\]

Gibbons and
Tsar\"ev~\cite{GT} were first who noticed  that the chordal
L\"owner equation plays an essential role in the classification of
reductions of the Benney equations. If we denote
\[
u=u_0=A^0,\dots,  u_N=A^N,\quad A^n=A^n(\ub),\quad n>N,
\]
then we obtain $N(N-1)/2$ compatibility conditions for $A^{N+1}$ and the function $z=z(\lambda,\tb)$ inverse to $\lambda$ satisfies the equations
\begin{equation}\label{p1}
\partial_k z=-\frac{\partial_ku}{z-\mu_k},
\end{equation}
where $\partial_k=\frac{\partial}{\partial r^k}$ and $\mu_k$ are the zeros of the function $\lambda_z$ and the values $r^k=\lambda(\mu_k,\tb)$ are the Riemann invariants. Formally, the equation above is similar to the chordal
L\"owner equation which we will discuss in what follows. 
The consistency conditions for \eqref{p1} are
\begin{equation}\label{p2}
\partial_i \mu_k=\frac{\partial_k u}{\mu_i-\mu_k},\quad \partial_i\partial_k u=2\frac{\partial_i u \partial_k u}{(\mu_i-\mu_k)^2},
\end{equation}
which is the Gibbons--Tsar\"ev system.

A generalization was suggested by Ma{\~n}as, Mart{\'\i}nez Alonso and Medina \cite{MMM}, where
the authors were looking for the function $\lambda$ and its inverse as a solution to the system of equation
\[
\frac{\partial z}{\partial u_k}=\sum\limits_{k=1}^N\frac{\eta_{ik}}{z-\mu_k},
\]
satisfying some compatibility conditions. Formally, again the above equations are of the form of multi-slit chordal L\"owner equation.

Later Takebe, Teo, and
Zabrodin~\cite{Takebe06} showed that the chordal and radial
L\"owner PDE served as consistency conditions for one-variable
reductions of dispersionless KP and Toda hierarchies,
respectively. In the chordal case, the function $\lambda$ satisfying \eqref{dKPE} depends on $\tb$ via one function $s(\tb)$ and
\[
\frac{\partial \lambda}{\partial s}=-\frac{k}{z-\xi}\frac{\partial \lambda}{\partial z},
\]
with the compatibility condition of hydrodynamic type
\[
\frac{\partial s}{\partial t_n}=\chi_n \frac{\partial s}{\partial x},
\]
where $k$ is the $s$-derivative of the coefficient at $1/z$ in the Laurent expansion of $\lambda$, the functions $\chi_n(s)$ are constructed in a canonical way from the Lax function, and again we see the chordal L\"owner PDE. These approaches are somewhat in want of analytic background of the L\"owner theory.

On the other hand, another evolution process described by
Laplacian growth~\cite{GV} possesses an infinite number of
conservation laws, harmonic moments. Being a typical field problem
the moments of the Laplacian growth bring us to the dispersionless
Toda hierarchy~\cite{Mineev}. Unlike the Laplacian growth the
L\"owner evolution represents another group of models, in which
the evolution is governed by the infinite number of parameters,
namely the controllable dynamical system, where the infinite number
of degrees of freedom follows from the infinite number of driving
terms. Surprisingly, the same structural background, the Virasoro
algebra, appears again for this group~\cite{MarkinaVasiliev10}.

The idea of this paper is to revisit Gibbons and Tsar\"ev
observation and show that the chordal L\"owner evolution also
possesses  an infinite number of conservation laws, moments. We
show that the L\"owner PDE is exactly the Vlasov equation under an
appropriate change of variables and the L\"owner ODE implies the
hydrodynamic type conservation equation. We start with the L\"owner evolution and
splitting time we arrive at integrable chains. This approach shows universality 
of the L\"owner equation as an attraction point for several integrable chains, this was noticed in \cite{Pavlov2006}.

\section{Vlasov and L\"owner equations, conservation laws}

Let us consider a L\"owner chain of receding domains $\mathbb
H_t=\mathbb H\setminus\gamma_t$ in the upper half-plane $\mathbb
H=\{:\im z>0\}$ and let $f\colon \mathbb H\to\mathbb H_t$ is
normalized near infinity as
\begin{equation}\label{norm}
f(z,t)=z+\frac{A^0}{z}+O\left(\frac{1}{z^2}\right),
\end{equation}
where $(-A^0(t))$ is the half-plane capacity of $\gamma_t$. Let
$\gamma_t$ be a Jordan curve in $\mathbb H$ except for an end
point on the real axis $\mathbb R$, $\gamma_t$ is parameterized by
$t$. Then $f$ satisfies the L\"owner PDE
\begin{equation}\label{LPDE}
(z-\xi_t)\frac{\partial f(z,t)}{\partial t}-\frac{d A^0}{dt}\frac{\partial f(z,t)}{\partial z}=0,
\end{equation}
with a real-valued continuous driving function $\xi_t$ and an
initial condition $f(z,0)=f_0(z)$. For every $t\geq0$, the
function $f(z,t)$ has a continuous extension on the closure of
$\mathbb H$, and the extended function denoted also by $f(z,t)$
satisfies equation \eqref{LPDE} at least almost everywhere. The
driving function $\xi_t$ generates the growing slit $\gamma_t$.

 %%%%%%%%%%%%%%%%%%%%%%%%Phase%%%%%%%%%%%%%%%%%%%%%%%%%%%%%%
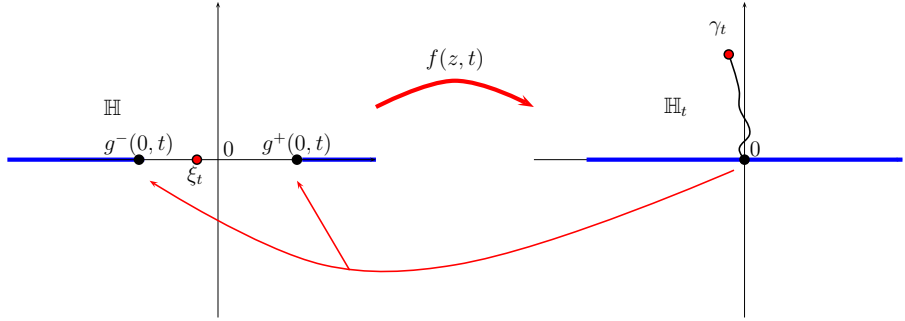
\begin{figure}[ht] \scalebox{0.7}{
\begin{pspicture}(2,1)(17,8)
%\psframe[linecolor=lightgray, fillstyle=solid,
%fillcolor=lightgray](11,4)(18,7.5)
\psline[linecolor=blue,linewidth=0.8mm](1,4)(3.6,4)
\psline[linecolor=blue,linewidth=0.8mm](6.6,4)(8,4)
%\rput(15.3,7){$y$}\rput(18,4.3){$x$}
\rput(13.7,5){$\mathbb H_t$}\rput(15.2,4.2){0}
\psline[linewidth=0.15mm]{->}(15,1)(15,7)
\psline[linewidth=0.15mm]{->}(11,4)(18,4)
%%%%%
%\psframe[linecolor=lightgray, fillstyle=solid,
%fillcolor=lightgray](1,4)(8,7.5)
%\rput(4.65,5.25){\includegraphics[height=1 in]{RandomCurve.eps}}
 \pscurve[linewidth=0.3mm](15,4)(14.9,4.2)(15.1,4.5)(14.9,5)(14.9,5.3)(14.7,6)
\psline[linecolor=blue,linewidth=0.8mm](12,4)(18,4)
%\rput(5.3,7){$\eta$}\rput(8,4.3){$\xi$}
\rput(3,5){$\mathbb H$}\rput(5.2,4.2){0}
\psline[linewidth=0.15mm]{->}(5,1)(5,7)
\psline[linewidth=0.15mm]{->}(2,4)(8,4)
%\pscurve[linecolor=blue,linewidth=0.8mm](5,4)(4.9,4.3)(5.1,5)(5.5,5.4)(5.1,5.6)
\rput(14.5,6.5){$\gamma_t$}
 \pscurve[linewidth=0.8mm,
linecolor=red]{->}(8,5)(9.5,5.5)(11,5)
\rput(9.5,5.9){$f(z,t)$}
\pscurve[linewidth=0.3mm,
linecolor=red]{->}(14.8,3.8)(9.5,2)(7,2)(3.7,3.6)
\psline[linecolor=red,linewidth=0.3mm]{->}(7.5,1.9)(6.5,3.6)
\rput(3.5,4.3){$g^-(0,t)$}
\rput(6.5,4.3){$g^+(0,t)$}
\rput(4.57,3.7){$\xi_t$}
 \pscircle[fillstyle=solid,
fillcolor=black](3.5,4){.1}
 \pscircle[fillstyle=solid,
fillcolor=red](14.7,6){.1}
 \pscircle[fillstyle=solid,
fillcolor=red](4.6,4){.1}
 \pscircle[fillstyle=solid,
fillcolor=black](6.5,4){.1}
 \pscircle[fillstyle=solid,
fillcolor=black](15,4){.1}
\end{pspicture}}
%\caption[]{The mapping $h(z,t)$}
\end{figure}
%%%%%%%%%%%%%%%%%%%%%%%%%%%%%%%%%%%%%%%%%%%%%%%%%%%%%%%%%%%%%%

We will also use the two-parametric family of conformal maps
$$g(w,t,\tau):=f^{-1}(w(z,t),\tau)=f^{-1}(f(z,t),\tau),$$ where $0\leq \tau\leq
t<\infty.$ We also denote $g(w,t,0)=:g(w,t)$. The function $g$
maps the half-plane $\mathbb H$ onto a subset of $\mathbb H$.  It
satisfies the L\"owner ODE for the half-plane
\begin{equation}\label{LODE}
\frac{\partial g(w,t,\tau)}{\partial t}=-\frac{d A^0/d
t}{g(w,t,\tau)-\xi_t}, \quad 0\leq \tau\leq t<\infty,\quad
g(w,\tau,\tau)=w.
\end{equation}
Moreover,
$\lim\limits_{t\to\infty}g(w,t,\tau)=\lim\limits_{t\to\infty}f^{-1}(f(z,t),\tau)=f(z,\tau)$.

Define the time splitting as the real-valued functions $t=t(x,s)$, a solution to the
quasi-linear differential equation
\begin{equation}\label{hydro}
\xi_t\frac{\partial t}{\partial x}+\frac{\partial t}{\partial s}=0,
\end{equation}
satisfying the asymptotic behaviour
$\lim\limits_{x\to\infty}t(x,s)=\lim\limits_{x\to-\infty}t(x,s)<\infty$.
Assume that $\xi_t$ is a function which admits a cone of solutions
to \eqref{hydro} with the needed asymptotic behaviour.

Now, let us consider the superposition $f(z,t(x,s))$ and multiply both sides of \eqref{LPDE} by
$\frac{\partial t}{\partial x}$. By abuse of notation, we continue to write $f$  for the function
$f(z,t(x,s))=f(z,x,s)$. Then
\[
z\frac{\partial f}{\partial x}-\xi_t\frac{\partial f(z,t)}{\partial t}\frac{\partial t}{\partial x}-\frac{\partial A^0}
{\partial x}\frac{\partial f}{\partial z}=0.
\]
If we use equation \eqref{hydro}, then
\begin{equation}\label{Vlasov}
z\frac{\partial f}{\partial x}+\frac{\partial f}{\partial s}-\frac{\partial A^0}{\partial x}\frac{\partial f}
{\partial z}=0,
\end{equation}
which is the Vlasov equation, see~\cite{GT, Vlasov61},  describing time evolution of the distribution function of plasma
consisting of charged particles with long-range  interaction.
In fluid descriptions of plasmas  one does not consider the velocity distribution but rather the plasma moments
$A^n(t(x,s))\equiv A^n(x,s)$.

Among solutions to the Vlasov equation \eqref{Vlasov} let us
choose those which provide finite integrals for the moments $A^n$.
Namely, for a given solution $f(z,x,s)$ with the normalization
\eqref{norm}, choose a solution $\phi(z,x,s)=\varphi(f(z,x,s))$
where $\varphi$ is an appropriate rapidly decreasing at infinities
$z\to\pm\infty$ function, see, e.g., \cite{PavlovTsarev2013}.
For example, $\varphi(f)=\exp(-f^2)$ is appropriate. Then the moments
$A^n(x,s)$ are defined by
\[
A^n(x,s)=\int_{-\infty}^{\infty}w^n\,\phi(w,x,s)\,dw,\quad n\geq 1.
\]
The direct computations implies
\[
A^n_s=\int_{-\infty}^{\infty}w^n\,\frac{\partial \phi}{\partial
s}\,dw,\quad A^{n+1}_x=\int_{-\infty}^{\infty}w^{n+1}\,
\frac{\partial \phi}{\partial x}\,dw.
\]
Integrating by parts yields
\[
A^{n-1}=-\int_{-\infty}^{\infty}\frac{w^n}{n}\,\frac{\partial \phi
}{\partial w}\,dw.
\]
Now we can use the Vlasov equation \eqref{Vlasov}  and arrive at the equation for the moments
\begin{equation}\label{mom}
A^n_s+A^{n+1}_x+nA^{n-1}A^0_x=0,
\end{equation}
which is an infinite autonomous system, known as Benney's moment equations, see~\cite{Benney}, which appear
in long wavelength hydrodynamics of an ideal incompressible fluid of a finite depth in a gravitational field.

Following \cite{GT, KM} let us define a function $\lambda(z,x,s)$ by the Cauchy principal value of a singular integral
\[
\lambda(z,x,s)=z+\int_{-\infty}^{\infty}\frac{\phi(w,x,s)}{z-w}dw=z+\sum_{n=0}^{\infty}\frac{A^n}{z^{n+1}},\quad
\text{ $z\to \infty$ in $\mathbb H$},
\]
where $z=g(w,t(x,s))$ and the coefficient $A^0$ is the same as for $f$. Then,
\[
\lambda_s=\frac{\partial \lambda}{\partial s}=z_s+\sum_{n=0}^{\infty}\left(\frac{A^n_s}{z^{n+1}}-\frac{(n+1)A^nz_s}
{z^{n+2}}\right),
\]
\[
\lambda_x=\frac{\partial \lambda}{\partial x}=z_x+\sum_{n=0}^{\infty}\left(\frac{A^n_x}{z^{n+1}}-\frac{(n+1)A^nz_x}
{z^{n+2}}\right),
\]
and
\[
\lambda_s+z\lambda_x=z_s+z\cdot z_x+A^0_x+\sum_{n=0}^{\infty}\frac{A^n_s+A^{n+1}_x-nA^{n-1}z_s-(n+1)A^nz_x}{z^{n+1}}.
\]
Making use of the moment equations we come to
\[
\lambda_s+z\lambda_x=z_s+z\cdot z_x+A^0_x+\sum_{n=0}^{\infty}\frac{-nA^{n-1}A^0_x}{z^{n+1}}-\sum_{n=1}^{\infty}
\frac{nA^{n-1}(z_s+z\cdot z_x)}{z^{n+1}}=
\]
\[
=A^0_x\lambda_z+(z_s+z\cdot z_x)\left(1-\sum_{n=1}^{\infty}\frac{nA^{n-1}}{z^{n+1}}\right)=
\lambda_z\left(A^0_x+z_s+z\cdot z_x\right).
\]
The L\"owner ODE \eqref{LODE} implies that $A^0_x=z_x(\xi_t-z)$ and the definition of the function $t(x,s)$ yields that
\begin{equation}\label{GibbonsEq}
A^0_x+z_s+z\cdot z_x=0,
\end{equation}
and therefore,
the equality $\lambda_s+z\lambda_x=0$ holds along the trajectories of the L\"owner ODE \eqref{LODE}. Equation
\eqref{GibbonsEq} received the name the Gibbons equation in \cite{Pavlov2007} following the original Gibbons' paper
\cite{Gibbons1982}.

Let us consider the map $z(\lambda,x,s)$ which is the inverse to $\lambda(z,x,s)$ with respect to
$\lambda \leftrightarrow z$,
\begin{eqnarray}
\label{ser1} \lambda(z,x,s)&=&z+\sum_{n=0}^{\infty}\frac{A^n}{z^{n+1}},\quad \text{ $z\to \infty$ in $\mathbb H$},\\
\label{ser2} z(\lambda, x,s)&=&\lambda-\sum_{n=0}^{\infty}\frac{H^n}{\lambda^{n+1}}.
\end{eqnarray}
Then,
\[
\sum_{n=0}^{\infty}\frac{A^n}{z^{n+1}}=\sum_{n=0}^{\infty}\frac{H^n}{\lambda^{n+1}},
\]
and
\[
\frac{\lambda}{z}\left(A^0+\frac{A^1}{z}+\dots\right)=H^0+\frac{H^1}{\lambda}+\dots
\]
So $H^0=A^0$. We continue by
\[
\lambda\left(\frac{\lambda}{z}-1\right)A^0+\frac{\lambda^2}{z^2}\left(A^1+\frac{A^2}{z}+\dots\right)=
H^1+\frac{H^2}{\lambda}+\dots,
\]
and conclude $H^1=A^1$. In the same fashion we come to
\[
\lambda^2\left(\frac{\lambda}{z}-1\right)A^0+\lambda\left(\frac{\lambda^2}{z^2}-1\right)A^1+
\frac{\lambda^3}{z^3}\left(A^2+\frac{A^3}{z}+\dots\right)=H^2+\frac{H^3}{\lambda}+\dots,
\]
and $H^2=A^2+(A^0)^2$.
Finally, we have
\[
\sum_{k=0}^{n}\lambda^{n-k}\left(\frac{\lambda^{k+1}}{z^{k+1}}A^k-H^k\right)+\frac{\lambda^{n+1}}{z^{n+2}}
\left(A^{n+1}+\frac{A^{n+2}}{z}+\dots\right)=\frac{1}{\lambda}\left(H^{n+1}+\frac{H^{n+2}}{\lambda}+\dots\right),
\]
and the coefficient $H^n$ is calculated as $H^n=A^n+P(A^0,\dots, A^{n-1})$, where $P(A^0,\dots, A^{n-1})$ is a
polynomial of $A^0,\dots, A^{n-1}$, $n\geq 2$.
The first coefficients are
\[
H^0=A^0,\quad H^1=A^1,\quad H^2=A^2+(A^0)^2, \quad H^3=A^3+3A^0A^1,
\]
\[
H^4=A^4+4A^0A^2+2(A^1)^2+2(A^0)^3.
\]
Analogous coefficients were calculated in, e.g., \cite{KM, Tammi}.

This way the  L\"owner ODE \eqref{LODE} becomes the conservation
equation in the following sense. According to \eqref{GibbonsEq}
\[
\frac{d}{d s}\int_{-\infty}^{\infty}z(\lambda, x,
s)\,dx=-\int_{-\infty}^{\infty} (A^0_x+z\cdot z_x)\,dx,
\]
where we integrate with respect to $x\in\mathbb R$ in the Cauchy
principal value sense. The requirements on the asymptotic
behaviour of $t(x,s)$ as $x\to\pm\infty$ imply that
\[
\int_{-\infty}^{\infty} (A^0_x+z\cdot z_x)\,dx=0,
\]
Therefore
\[
\frac{d}{d s}\int_{-\infty}^{\infty}z(\lambda, x, s)\,dx=0,
\]
which corresponds to the momentum conservation law. So the
conserved quantities of the evolution are the moments
\[
I^n=\int_{-\infty}^{\infty}H^n(x, s)\,dx,\quad n\geq 0.
\]
Analogous integrals of motion were studied in the original work by
Benney~\cite{Benney} as well as in \cite{KM, PT, Zakharov}.

The Poisson structure allows us to reformulate the Benney moment
equation \eqref{mom} as an evolution equation with a Hamiltoinian
function. The Kupershmidt-Manin Poisson structure \cite{KM, KM2}
starts with the operators of differentiation and multiplication to
the right  for the moments $A^n\frac{\partial}{\partial x}$ as
skew-symmetric operators with respect to the $L^2(\mathbb
R)$-paring, acting to the right by
\[
\{A^m,A^n\}(\cdot)=-mA^{n+m-1}\frac{\partial}{\partial x}(\cdot)-n\frac{\partial}{\partial x}
\left(A^{n+m-1}(\cdot)\right).
\]
Then for any two observables $F(A)$ and $G(A)$, the Poisson bracket can be written as
\[
\{F,G\}(A)=\sum\limits_{m,n=0}^{\infty}\int\limits_{-\infty}^{\infty}\frac{\delta F}{\delta A^m}\{A^m,A^n\}
\frac{\delta G}{\delta A^n}dx.
\]
Writing $\bar{H}^k=\frac{1}{k}\int\limits_{-\infty}^{\infty}
H^kdx=\frac{1}{k}I^k$, we have the hierarchy of commuting flows
with the Hamiltonians $\bar{H}^k\colon \{\bar{H}^k,\bar{H}^j\}=0$,
in the form of evolution equations
\[
\frac{\partial A^{m}}{\partial t_k}=\sum\limits_{n=0}^{\infty}\{A^m,
A^n\}\frac{\delta \bar{H}^k}{\delta A^n},
\]
so that equation \eqref{mom} becomes the second equation in this hierarchy.

\section{Finite-dimensional time}

The L\"owner PDE \eqref{LPDE} can be generalized to the form
\begin{equation}\label{mLPDE}
\frac{\partial f(z,t)}{\partial
t}=\sum_{k=1}^m\frac{\mu_k(t)}{z-\xi_k(t)}\frac{d
A^0}{dt}\frac{\partial f(z,t)}{\partial z},\;\;\;f(z,0)=f_0(z),
\end{equation}
with piecewise continuous coefficients $\mu_k(t)\geq0$,
$k=1,\dots,m$, $\sum_{k=1}^m\mu_k(t)=1$, and real-valued
continuous driving functions $\xi_1(t),\dots,\xi_m(t)$. A solution
$f(z,t)$ to \eqref{mLPDE} maps $\mathbb H$ onto $\mathbb
H\setminus\cup_{k=1}^m\gamma_k(t)$ where $\gamma_k(t)$ are growing Jordan
curves (slits) in $\mathbb H$ except for their endpoints on $\mathbb R$.
The driving functions $\xi_k(t)$ generate slits $\gamma_k(t)$, and the
coefficients $\mu_k(t)$ govern the relative dynamics of the slits
$\gamma_k(t)$ with respect to each other.

Instead of \eqref{mLPDE}, it is possible to consider the
generalized L\"owner PDE with a generalized time-vector ${\bf t}=(t_1,\dots,t_m)$
\begin{equation}\label{vecLPDE}
(z-\xi_k({\bf t}))\frac{\partial f(z,{\bf t})}{\partial
t_k}=\frac{\partial A^0}{\partial t_k}\frac{\partial f(z,{\bf
t})}{\partial z},\;\;\;f(z,{\bf 0})=f_{\bf
0}(z),\;\;\;k=1,\dots,m.
\end{equation}

In this model, $A^0({\bf t})=A^0(t_1,\dots,t_m)$ is not an
arbitrary function of $\mathbf{t}$. At every point ${\bf t}=(t_1,\dots,t_m)$,
\[
\frac{\partial A^0}{\partial t_1}=\frac{\partial A^0}{\partial
t_2}=\dots=\frac{\partial A^0}{\partial t_m}.
\]

For every ${\bf t}=(t_1,\dots,t_m)$, the solution $f(z,{\bf t})$ to
 system \eqref{vecLPDE} maps $\mathbb H$ onto $\mathbb
H\setminus\cup_{k=1}^m\gamma_k(t_k)$, where $\gamma_k(t_k)$ is an endpoint of the slit $\gamma_k$ generated by the driving function
$\xi_k({\bf t})$.

Similarly to the scalar $t$, let us denote by
\[
g(w,\tau,{\bf t}):=f^{-1}(w(z,\tau),{\bf t})=f^{-1}(f(z,\tau),{\bf
t}),\;\;\; \tau=(\tau_1,\dots,\tau_m),
\]
where $0\leq\tau_k\leq t_k<\infty$ for any $k=1,\dots,m.$ We also
write $g(w,{\bf 0},{\bf t})=:g(w,{\bf t})$, where $\mathbf{0}$ states for the null-vector. The function $g$ maps
the half-plane $\mathbb H$ onto a subset of $\mathbb H$. It
satisfies the system of L\"owner's ODE in the half-plane
\begin{equation}\label{vecLODE}
\frac{\partial g(w,\tau,{\bf t})}{\partial t_k}=-\frac{\partial
A^0/\partial t_k}{g(w,\tau,{\bf t})-\xi_k({\bf t})}, \quad 0\leq
\tau_j\leq t_j<\infty, \quad g(w,\tau,\tau)=w,
\end{equation}
\[
j=1,\dots,m,\;\;\;k=1,\dots,m.
\]
Moreover,
$\lim\limits_{t_k\to\infty}g(w,\tau,(\tau_1,\dots,\tau_{k-1},t_k,\tau_{k+1},\dots,\tau_m))=
f(z,\tau)$.

Again we can define the vector-function ${\bf t}={\bf t}(x,s)$, as
a solution to the system of quasi-linear differential equations
\begin{equation}\label{vechydro}
\xi_k({\bf t})\frac{\partial t_k}{\partial x}+\frac{\partial
t_k}{\partial s}=0,\;\;\;k=1,\dots,m,
\end{equation}
satisfying the asymptotic behaviour
$\lim\limits_{x\to\infty}t_k(x,s)=\lim\limits_{x\to-\infty}t_k(x,s)<\infty$,
$k=1,\dots,m.$ Assume that functions $\xi_k(t_k)$ admit their
cones of solutions to \eqref{vechydro} under the necessary asymptotic
behaviour.

Equation \eqref{vecLPDE} implies that $f(z,{\bf
t}(x,s))=:f(z,x,s)$ satisfies
\[
z\frac{\partial f}{\partial t_k}\frac{\partial t_k}{\partial
x}-\xi_k({\bf t})\frac{\partial f}{\partial t_k}\frac{\partial
t_k}{\partial x}-\frac{\partial A^0}{\partial t_k}\frac{\partial
t_k}{\partial x}\frac{\partial f}{\partial z}=0
\]
which together with \eqref{vechydro} gives
\[
z\frac{\partial f}{\partial t_k}\frac{\partial t_k}{\partial
x}+\frac{\partial f}{\partial t_k}\frac{\partial t_k}{\partial
s}-\frac{\partial A^0}{\partial t_k}\frac{\partial t_k}{\partial
x}\frac{\partial f}{\partial z}=0,\;\;\;k=1,\dots,m.
\]

Summing up the latter equations for $k=1,\dots, m$, we obtain the
Vlasov equation for $f(z,{\bf t}(x,s))$
\[
z\frac{\partial f(z,{\bf t}(x,s))}{\partial x}+\frac{\partial
f(z,{\bf t}(x,s))}{\partial s}-\frac{\partial A^0}{\partial
x}\frac{\partial f(z,{\bf t}(x,s))}{\partial z}=0.
\]

Similarly to the scalar case, there appear the moments $A^n({\bf
t}(x,s))=:A^n(x,s)$ satisfying equation \eqref{mom}, the functions
$\lambda(z,x,s)$ and $z(\lambda,x,s)$ and the hierarchy of
commuting flows with the Hamiltonians $\bar H^k$.

Equations \eqref{vecLPDE} can be reduced to \eqref{mLPDE}. Indeed,
assume that $\mu_1(t)>0$ and construct the following reduction
\[
\frac{dt_k}{dt_1}=\frac{\mu_k}{\mu_1},\;\;\;t_k(0)=0,\;\;\;k=2,\dots,m.
\]

Then, after multiplying by $\frac{\partial t_k}{\partial t_1}$ and
summing up, equations \eqref{vecLPDE} for $t_1=t$ and $f(z,{\bf
t}(t))=:f(z,t)$ become
\[
\frac{\partial f(z,{\bf t}(t))}{\partial t}=\frac{\partial
f}{\partial t_1}+\frac{\partial f}{\partial t_2}\frac{\partial
t_2}{\partial t}+\dots+\frac{\partial f}{\partial
t_m}\frac{\partial t_m}{\partial t}=
\frac{1}{\mu_1}\left[\frac{\mu_1}{z-\xi_1}\frac{\partial
A^0}{\partial t_1}+\dots+\frac{\mu_m}{z-\xi_m}\frac{\partial
A^0}{\partial t_m}\right]\frac{\partial f}{\partial z}
\]
which is equivalent to \eqref{mLPDE} with $\tilde A^0=A^0/\mu_1$.

\section{Infinite-dimensional time}

The limiting case of equation \eqref{mLPDE} as $m\to\infty$ leads to
the L\"owner-Kufarev type equation
\begin{equation}\label{count}
\frac{\partial f}{\partial t}=\int_{\mathbb
R}\frac{d\nu_t(\xi)}{z-\xi(t)}\;\frac{dA^0}{dt}\;\frac{\partial
f}{\partial z},
\end{equation}
where, for every $t\geq0$, $d\nu_t(\xi)$ is a probability measure
with a compact support $I_t\subset\mathbb R$. A solution $f(z,t)$
to \eqref{count} maps $\mathbb H$ onto $\mathbb H\setminus K_t$
where, in general, the omitted set $K_t\cap\mathbb H$ is not
reduced to a countable set of slits. The set $K_t$ is generated by
the measure $d\nu_t(\xi)$.

However, the domain $\mathbb H\setminus K_t$ is the Carath\'eodory
kernel for the sequence of domains $\mathbb
H\setminus\cup_{k=1}^m\gamma_k(t)$ as $m\to\infty$. Here the slits
$\gamma_k$ are dense in $K_t$. In this interpretation the measure
$d\nu_t(\xi)$ is represented as a limit of point mass measures
with a dense set of mass points in the support of $d\nu_t$.

In this case it is impossible to generalize directly system
\eqref{vecLPDE} passing to an infinite set of equations
corresponding to the countable set of coordinates $(t_1,t_2,\dots)$.
Let us build a model with a successive dynamics of every slit
$\gamma_1,\gamma_2,\dots$. For $k=1,2,\dots$, denote by
\[
P_k(z,t_k)=\frac{\partial A^0}{\partial t_k}\frac{\partial
f(z,t_k)}{\partial z}\;\;\text{for}\;\;T_{k-1}<t_k<T_k,\;\;T_0=0,
\]
and
\[
P_k(z,t_k)=0\;\;\text{for}\;\;t_k\in\mathbb
R\setminus(T_{k-k},T_k),\;\;\;k=1,2,\dots\;.
\]

Now, instead of \eqref{count}, we are able to introduce a system
of PDE with an infinite set of coordinates ${\bf
t}:=(t_1,t_2,\dots)$,
\begin{equation}\label{infLPDE}
(z-\xi_k(t_k))\frac{\partial f}{\partial
t_k}=P_k(z,t_k),\;\;\;k=1,2,\dots\;.
\end{equation}

Suppose a function $f_0(z)$ is expanded near infinity as
\begin{equation}\label{norm2}
f_0(z)=z+\sum_{n=0}^{\infty}\frac{A^n}{z^{n+1}},
\end{equation}
and let $f_0$ serve as an initial data for the L\"owner chain $f(z,t)$
governed by \eqref{count} and as an initial data for the first equation of system \eqref{infLPDE}. Successively, the function $f(z,t_k)$  serves as an initial data for the $(k+1)$-th equation in \eqref{infLPDE}. It is clear that the resulting chain
$f(z,t)=f(z,{\bf t})$, ${\bf t}=(t_1,t_2,\dots)$, in contrast to the chain obtained from~\eqref{count}, is piecewise differentiable.  The functions $f(z,t)$ are normalized
as in \eqref{norm2} with $A^n=A^n(t)$. So there exist driving
functions $\xi_1(t_1),\xi_2(t_2),\dots$ such that $f(z,{\bf t})$,
${\bf t}=(t_1,t_2,\dots)$, is a solution to the infinite system of
PDE \eqref{infLPDE}.

Let us apply the results by Takebe, Teo and Zabrodin
\cite{Takebe06} to construct a one-variable reduction of
dispersionless KP hierarchy for the system of PDE \eqref{infLPDE}.

Let $g(w,{\bf t}):=f^{-1}(w,{\bf t})$ be the inverse to $f(z,{\bf
t})$. Then  $g$ is normalized at infinity as
\[
g(w,{\bf t})=w+\sum_{n=1}^{\infty}\frac{b_n({\bf t})}{w^n}.
\]
Denote by $\Phi_k(w,{\bf t})=[g^k(w,{\bf t})]_{\geq 0}$, $k\geq1$, the Faber polynomials
for $g(w,{\bf t})$. Let us forget for the moment the dependence on ${\bf t}$ and let us  write simply $g(w)$ and  $\Phi_k(w)$. The first Faber polynomials are
\[
\Phi_0=1,\quad \Phi_1=w,\quad \Phi_2=w^2-2b_1,\quad \Phi_3=w^3-3b_1w-3b_2,
\]
and the recurrence formula 
\[
\Phi_{n+1}=w\Phi_n-\sum\limits_{k=1}^{n-1}b_{n-k}\Phi_k-(n+1)b_n
\]
holds for all $n\geq 1$. The Faber polynomials are related to the Grunsky coefficients which implies that
\[
\log\frac{g(w)-\xi}{w}=-\sum\limits_{n=1}^{\infty}\frac{1}{nw^n}\Phi_n(\xi).
\]  
Changing variables $\xi=e^u$ and differentiating both sides with respect to $u$ yields
\[
\frac{1}{g(w)-\xi}=\sum\limits_{n=1}^{\infty}\frac{1}{nw^n}\Phi'_n(\xi).
\]
Returning back to equation~(\ref{infLPDE}) we conclude that the
function $g=f^{-1}$, w.r.t. the first variable,  satisfies the system of equations
\[
\frac{\partial g}{\partial t_k}=-\frac{\partial A^0}{\partial
t_k}\sum_{n=1}^{\infty}\frac{\Phi_n'(\xi_k,t_k)}{nw^n},\;\;\;T_{k-1}<t_k<T_k,
\]
and
\[
\frac{\partial g}{\partial t_k}=0\;\;\text{for}\;\;t_k\in\mathbb
R\setminus(T_{k-1},T_k),\;\;\;k=1,2,\dots\;.
\]
This implies that, for all $k\geq1$,
\[
n\frac{\partial b_k}{\partial t_k}=-\frac{\partial A^0}{\partial
t_k}\Phi_k'(\xi_k,t_k),\;\;\;T_{k-1}<t_k<T_k,
\]
and
\[
\frac{\partial b_k}{\partial t_k}=0,\;\;\;t_k\notin(T_{k-1},T_k).
\]

There is an evident way to write dependence on ${\bf t}$ through a
single variable $t=t_1$. Set
\[
\tau({\bf t})=t_1,\;\;\text{if}\;\;t_1\in(0,T_1),
\]
\[
\tau({\bf
t})=kt_k(t_1)\;\;\text{if}\;\;t_1\in(T_{k-1},T_k),\;\;k\geq2,
\]
where
\[
\frac{dt_k}{dt_1}=\Phi_k'(\xi_k,t_k)\;\;\text{for}\;\;t_1\in(T_{k-1},T_k)
\]
and
\[
\frac{dt_k}{dt_1}=0\;\;\text{for}\;\;t_1\notin(T_{k-1},T_k),\;\;\;k\geq2.
\]

In the spirit of the results of Takebe, Teo and Zabrodin \cite{Takebe06} we conclude that the non-intersecting intervals $(T_{k-1},T_k)$ imply 
that given $f(z,\tau({\bf t}))$ as the solution to system \eqref{infLPDE} with the initial condition $f_0$, one
has the Lax function $\mathcal{L}=f(z,\tau({\bf t}))$ which solves the dKP hierarchy by
\[
\frac{\partial \mathcal L}{\partial t_k}=\{\mathcal L_k,\mathcal L\},\quad T_{k-1}<t_k<T_k,
\]
where $\mathcal L_k=\frac{1}{k}[ \mathcal L^k]_{\geq 0}$, and the Poisson bracket is given by
\[
\{F,G\}=\frac{\partial F}{\partial w}\frac{\partial G}{\partial x}-\frac{\partial F}{\partial x}\frac{\partial G}{\partial w}, \quad T_{k-1}<x:=t_1<T_k.
\]
The Benney equations again can be recovered as the second equation of dKP in the following way. Set $s=t_2$. Then,
\[
\frac{\partial \mathcal L}{\partial s}=\{(z^2+2A^0),\mathcal L\},\quad T_{1}<s,x<T_2,
\]
where
\[
\mathcal{L}=z+\sum_{n=0}^{\infty}\frac{A^n}{z^{n+1}}.
\]
Equating the coefficients in front of powers of $z$ leads to the equations~\eqref{mom}. The higher equations in the hierarchy lead
to interesting PDEs due to the conditions on commuting flows (compatibility conditions) \eqref{compa}
For example, if $n=2$ and $m=3$ imply the dKP equation \eqref{KZ} (Zabolotskaya-Khokhlov equation~\cite{ZabKhokh}) for $A^{0}$.

\section{Vlasov kinetic equation}

Let us return back to the consistency conditions \eqref{p2} (the Gibbons-Tsar\"ev system), where $u(\mathbf{r})$ is a conservation law density and $\mu_{k}(\mathbf{r}
) $ are the characteristic velocities of $N$ component hydrodynamic type system
\begin{equation}
r_{s}^{i}+\mu_{i}(\mathbf{r})r_{x}^{i}=0,\quad i=1,\dots, N,  \label{raz}
\end{equation}
written in the Riemann invariants. System \eqref{raz} is integrable by the  generalized hodograph method, see \cite{T85, T91},  
and  has infinitely many conservation laws and commuting flows
\begin{equation}
r_{y}^{i}+\lambda_{i}(\mathbf{r})r_{x}^{i}=0,  \label{dva}
\end{equation}
where $\lambda_{i}$ are functions of two entries $\mu_{i}$ and $u$ only: $
\lambda_{i}=F(\mu_{i},u)$. We will make use of the Tsar\"ev system
\begin{equation}
\frac{\partial _{i}\lambda_{k}}{\lambda_{i}-\lambda_{k}}=\frac{\partial
_{i}\mu_{k}}{\mu_{i}-\mu_{k}},\text{ \ }i\neq k,  \label{tsar}
\end{equation}%
which is a direct consequence of the commutation $(r_{t}^{i})_{y}=(r_{y}^{i})_{t}$. Substitution of the ansatz $\lambda_{i}=F(\mu_{i},u)$ into (\ref{tsar}) yields
\begin{equation}
(\mu_{k}-\mu_{i})\frac{\partial F(\mu_{i},u)}{\partial u}=\frac{F(\mu_{k},u)-F(\mu_{i},u)}{\mu_{k}-\mu_{i}}-\frac{\partial F(\mu_{i},u)}{
\partial \mu_{i}},\text{ \ }i\neq k.  \label{a}
\end{equation}
Interchanging indices and summing up both formulas we obtain
\begin{equation}
(\mu_{k}-\mu_{i})\left( \frac{\partial F(\mu_{i},u)}{\partial u}+\frac{
\partial F(\mu_{k},u)}{\partial u}\right) =\frac{\partial F(\mu_{k},u)}{
\partial \mu_{k}}-\frac{\partial F(\mu_{i},u)}{\partial \mu_{i}}.
\label{b}
\end{equation}
In the limit $\mu_{k}\rightarrow \mu_{i}$ this formula becomes
\begin{equation}
2\frac{\partial F(\mu_{i},u)}{\partial u}=\frac{\partial ^{2}F(\mu_{i},u)}{
(\partial \mu_{i})^{2}}.  \label{j}
\end{equation}
Then (\ref{b}) reads
\begin{equation*}
(\mu_{k}-\mu_{i})\left( \frac{\partial ^{2}F(\mu_{i},u)}{(\partial \mu
^{i})^{2}}+\frac{\partial ^{2}F(\mu_{k},u)}{(\partial \mu_{k})^{2}}\right)
=2\left( \frac{\partial F(\mu_{k},u)}{\partial \mu_{k}}-\frac{\partial
F(\mu_{i},u)}{\partial \mu_{i}}\right) .
\end{equation*}
Taking derivative of this relationship with respect to $\mu_{k}$, we obtain
\begin{equation*}
\frac{\partial ^{3}F(\mu_{k},u)}{(\partial \mu_{k})^{3}}=\frac{\frac{%
\partial ^{2}F(\mu_{k},u)}{(\partial \mu_{k})^{2}}-\frac{\partial
^{2}F(\mu_{i},u)}{(\partial \mu_{i})^{2}}}{\mu_{k}-\mu_{i}},\text{ \ }%
i\neq k.
\end{equation*}
Interchanging indices we conclude that 
\begin{equation*}
\frac{\partial ^{3}F(\mu_{k},u)}{(\partial \mu_{k})^{3}}=a^{\prime }(u)
\end{equation*}
for any index $k$. Thus
\begin{equation}
F(\mu_{k},u)=\frac{1}{6}a(u)(\mu_{k})^{3}+b(u)(\mu_{k})^{2}+c(u)\mu_{k}+d(u),  \label{c}
\end{equation}
where functions $a(u),b(u),c(u),d(u)$ still have not been    determined yet.
However the
substitution (\ref{c}) into (\ref{j}) leads to $a(u)=$const$,b(u)=$const$
,2c^{\prime }(u)=a,d^{\prime }(u)=b$. So, (\ref{c}) admits the form
\begin{equation}
F(\mu_{k},u)=\frac{a}{6}(\mu_{k})^{3}+\frac{a}{2}u\mu_{k}+b[(\mu_{k})^{2}+u].  \label{k}
\end{equation}
Finally, the substitution (\ref{k}) into (\ref{a}) implies $a=0$. Thus we
have found the so called `dispersive relation'
\begin{equation}
\lambda_{i}=(\mu_{i})^{2}+u.  \label{tri}
\end{equation}

Following \cite{Fer} we write the L\"{o}wner system
\begin{equation}
\partial _{i}z=\frac{\partial _{i}u}{\mu_{i}-z}.  \label{zero}
\end{equation}
Then, see (\ref{raz}), (\ref{dva}), (\ref{tri}), (\ref{zero})),
\begin{equation*}
z_{x}=\sum \partial _{i}z\cdot r_{x}^{i}=\sum \frac{\partial _{i}u}{\mu
_{i}-z}r_{x}^{i},
\end{equation*}%
\begin{equation*}
-z_{s}=-\sum \partial _{i}z\cdot r_{s}^{i}=\sum \frac{\partial _{i}u}{\mu
_{i}-z}\mu _{i}r_{x}^{i}=\sum \partial _{i}u\cdot r_{x}^{i}+z\sum \frac{%
\partial _{i}u}{\mu _{i}-z}r_{x}^{i}=u_{x}+zz_{x},
\end{equation*}%
\begin{equation*}
-z_{y}=-\sum \partial _{i}z\cdot r_{y}^{i}=\sum \frac{\partial _{i}u}{\mu
_{i}-z}\lambda_{i}r_{x}^{i}=\sum \frac{\partial _{i}u}{\mu_{i}-z}[(\mu
_{i})^{2}+u]r_{x}^{i}
\end{equation*}%
\begin{equation*}
=\sum \partial _{i}u\cdot (\mu_{i}-z)r_{x}^{i}+2z\sum \partial _{i}u\cdot
r_{x}^{i}+(z^{2}+u)\sum \frac{\partial _{i}u}{\mu_{i}-z}r_{x}^{i}
\end{equation*}%
\begin{equation*}
=(v_{x}-zu_{x})+2zu_{x}+(z^{2}+u)z_{x}=v_{x}+zu_{x}+uz_{x}+z^{2}z_{x},
\end{equation*}%
where we have introduced a new function $v(\mathbf{r})$ such that $\partial
_{i}v=\mu_{i}\partial _{i}u$. Indeed the potential function $v$ exists because  the compatibility
condition $\partial _{i}(\partial _{k}v)=\partial _{k}(\partial _{i}v)$
leads to the identity according to the Gibbons--Tsar\"ev system (\ref{p2}).
Thus
we reconstructed two equations
\begin{equation}
z_{s}+\left( \frac{z^{2}}{2}+u\right)_x=0,\text{ \ }z_{y}+\left( \frac{z^{3}%
}{3}+uz+v\right)_x=0.  \label{gen}
\end{equation}%
Their compatibility condition $(z_{s})_{y}=(z_{y})_{s}$ leads to the equations
\begin{equation}
v_{x}+u_{s}=0,\text{ \ }v_{s}+u_{y}+uu_{x}=0,  \label{dKP}
\end{equation}
which are equivalent to \eqref{KZ}.

Now we introduce the so called vertex equation%
\begin{equation}
\partial _{\tau (\zeta )}z(\lambda )=\partial _{x}\ln (z(\lambda )-z(\zeta
)),  \label{ierar}
\end{equation}%
where $z(\lambda )$ is just a short notation for $z(\lambda
;t_0,t_1,t_2,...)$. Here we use infinitely many 
`time' variables $t_k$, which will be discussed below.
Let us consider the formal expansion \eqref{ser2} as $\zeta \rightarrow \infty $ and %
\begin{equation}
\partial _{\tau (\zeta )}=-\frac{1}{\zeta }%
\partial _{t_0}-\frac{1}{\zeta ^{2}}\partial _{t_1}-\frac{1}{\zeta ^{3}}%
\partial _{t_2}-...  \label{f}
\end{equation}%
Then one can obtain infinitely many equations
\[
\partial_{t_n}z(\lambda)=\partial_x \frac{\Phi_{n+1}(z(\lambda))}{n+1},\quad n\geq 0,
\]
where $\Phi_n$ stands for the Faber polynomials (see Section 4), or for the first polynomials,
\[
\partial _{t_0}z(\lambda )=z_x(\lambda ),\text{ \ }\partial
_{t_1}z(\lambda )=\left( \frac{z^{2}(\lambda )}{2}+H^{0}\right)_x,
\]
\[
\partial _{t_2}z(\lambda )=\left( \frac{z^{3}(\lambda )}{3}%
+H^{0}z(\lambda )+H^{1}\right)_x,...
\]
 identifying $x=t_0,s=-t_1,y=-t_2$ as well as $u=H^{0},v=H^{1}$ (see (%
\ref{gen})). Substituting a similar formal expansion ($\lambda
\rightarrow \infty $)%
\begin{equation}
z(\lambda )=\lambda -\frac{H^{0}}{\lambda }-\frac{H^{1}}{\lambda ^{2}}-\frac{%
H^{2}}{\lambda ^{3}}-...  \label{d}
\end{equation}%
in these generating functions of conservation laws leads to infinitely
many local conservation laws. For instance%
\begin{equation}
(H^{k})_{s}+\left( H^{k+1}-\frac{1}{2}\underset{m=0}{\overset{k-1}{\sum }}%
H^{m}H^{k-1-m}\right) _{x}=0,\text{ }k=0,1,2,...  \label{one}
\end{equation}%
This is nothing but the Benney hydrodynamic chain \eqref{BE},
written in the conservative form where all conservation law densities $H^{k}$
are polynomials with respect to moments $A^{m}$ as in Section~2.

Alternative expansions ($\zeta \rightarrow 0$)%
\begin{equation}
z(\zeta )=H^{-1}+\zeta H^{-2}+\zeta ^{2}H^{-3}+...,\text{ \ }\partial _{\tau
(\zeta )}=\partial _{t_{-1}}+\zeta \partial _{t_{-2}}+\zeta ^{2}\partial
_{t_{-3}}+...  \label{z}
\end{equation}%
lead to another generating functions of conservation laws, for instance%
\begin{equation*}
\partial _{t_{-1}}z(\lambda )=\partial _{x}\ln (z(\lambda )-H^{-1}),\text{ \ 
}\partial _{t_{-2}}z(\lambda )=-\partial _{x}\frac{H^{-2}}{z(\lambda )-H^{-1}%
},...
\end{equation*}%
Substituting expansion (\ref{d}) and the expansion ($\lambda \rightarrow 0$)%
\begin{equation*}
z(\lambda )=H^{-1}+\lambda H^{-2}+\lambda ^{2}H^{-3}+...
\end{equation*}%
implies extra infinitely many conservation laws (cf. (\ref{one})). If for
instance, we substitute the above expansion in (\ref{gen}), 
two additional conservation laws%
\[
\partial_sH^{-1}=-\left( \frac{1}{2}(H^{-1})^{2}+H^{0}\right)_x=-\partial_x\frac{\Phi_2(H^{-1})}{2},
\]
\[
\partial_yH^{-1}=-\left( \frac{1}{3}(H^{-1})^{3}+H^{0}H^{-1}+H^{1}\right)_x=-\partial_x\frac{\Phi_3(H^{-1})}{3}
\]
follow. Infinitely many conservative laws (cf. (\ref{one}))%
\begin{equation*}
(H^{-1})_{s}+\left( H^{0}+\frac{1}{2}(H^{-1})^{2}\right) _{x}=0,
\end{equation*}%
\begin{equation*}
(H^{k})_{s}+\left( H^{k+1}-\frac{1}{2}\overset{k-1}{\underset{m=0}{\sum }}%
H^{m}H^{k-1-m}\right) _{x}=0,\text{ }k=0,1,...
\end{equation*}%
also can be written as the modified Benney hydrodynamic chain (see details in 
\cite{MaksBenney})%
\begin{equation}
B_{s}^{k}+B_{x}^{k+1}+\frac{1}{2}B^{0}B_{x}^{k}+\frac{k+1}{2}%
B^{k}B_{x}^{0}+kB^{k-1}\left( \frac{1}{2}B^{1}-\frac{1}{8}(B^{0})^{2}\right)
_{x}=0,  \label{1}
\end{equation}%
where $H^{-1}=B^{0},H^{0}=B^{1},H^{1}=B^{2}+B^{0}B^{1}-\frac{1}{12}%
(B^{0})^{3}$,... This Modified Benney hydrodynamic chain is related to
the modified dKP equation (cf. (\ref{dKP}))%
\begin{equation}
H_{s}^{-1}+\left( H^{0}+\frac{1}{2}(H^{-1})^{2}\right) _{x}=0,\text{ \ }%
H^{0}_s=H^{-1}_y+\left( H^{0}H^{-1}+\frac{1}{3}(H^{-1})^{3}\right) _{x},
\label{MdKP}
\end{equation}%
which can be obtained from the compatibility condition $(\tilde{z}_{s})_{y}=(%
\tilde{z}_{y})_{s}$, where%
\begin{equation}
\tilde{z}_{s}+\left( \frac{\tilde{z}^{2}}{2}+H^{-1}\tilde{z}\right) _{x}=0,%
\text{ \ }\tilde{z}_{y}+\left( \frac{\tilde{z}^{3}}{3}+H^{-1}\tilde{z}%
^{2}+(H^{0}+(H^{-1})^{2})\tilde{z}\right) _{x}=0.  \label{dist}
\end{equation}%
One can derive the modified L\"{o}wner equations%
\begin{equation*}
\partial _{i}\tilde{z}=\tilde{z}\frac{\partial _{i}H^{-1}}{\mu_{i}-\tilde{z}%
-H^{-1}},
\end{equation*}%
which are equivalent to the original L\"{o}wner equations (\ref{zero}) by 
substituting $z=\tilde{z}+H^{-1}$ and $(\mu_{i}-H^{-1})\partial
_{i}H^{-1}=\partial _{i}H^{0}$.

Both hydrodynamic chains have the same local Hamiltonian structure%
\begin{equation*}
A_{s}^{k}=-\frac{1}{2}[(k+m)A^{k+m-1}\partial _{x}+mA_{x}^{k+m-1}]\frac{\partial H^{2}}{%
\partial A^{m}},\text{ \ }
\end{equation*}
\begin{equation*}
B_{s}^{k}=-\frac{1}{2}[(k+m)B^{k+m-1}\partial
_{x}+mB_{x}^{k+m-1}]\frac{\partial H^{1}}{\partial B^{m}},
\end{equation*}%
where their Hamiltonian densities are%
\begin{equation*}
H^{2}=A^{2}+(A^{0})^{2},\text{ \ }H^{1}=A^1=B^{2}+B^{0}B^{1}-\frac{1}{12}(B^{0})^{3}.
\end{equation*}%
Since all moments $A^{k}$ can be expressed via moments $%
B^{0},B^{1},...,B^{k},B^{k+1}$, the Kupershmidt--Manin Poisson brackets%
\begin{equation*}
\{B^{k},B^{m}\}=[(k+m)B^{k+m-1}\partial _{x}+mB_{x}^{k+m-1}]\delta
(x-x^{\prime })
\end{equation*}%
can be recalculated via moments $A^{s}$. This means that the Benney
hydrodynamic chain has at least two local Hamiltonian structures (see details
in \cite{MaksBenney}).

As in the previous particular case, we are looking for $N$ components commuting the
hydrodynamic reductions%
\begin{equation*}
r_{\tau (\zeta )}^{i}=w^{i}(\mathbf{r},\zeta )r_{x}^{i}.
\end{equation*}%
Then (\ref{ierar}) reduces to the form%
\begin{equation*}
\partial _{i}z(\lambda )=\frac{\partial _{i}z(\zeta )}{1-[z(\lambda
)-z(\zeta )]w^{i}(\mathbf{r},\zeta )}.
\end{equation*}%
Taking into account (\ref{zero}), one can obtain%
\begin{equation*}
w^{i}(\mathbf{r},\zeta )=\frac{1}{\mu_{i}-z(\zeta )}.
\end{equation*}%
Using expansions (\ref{f}), (\ref{z}), one can expand the generating
function (with respect to parameter $\zeta $ at infinity and about zero,
respectively) of infinitely many higher commuting flows%
\begin{equation*}
r_{\tau (\zeta )}^{i}=\frac{1}{\mu_{i}-z(\zeta )}r_{x}^{i}.
\end{equation*}

\end{document}